\documentclass[3p,times,twocolumn]{elsarticle}

\usepackage{ecrc}

\volume{00}

\firstpage{1}

\journalname{Nuclear and Particle Physics Proceedings}

\runauth{}

\jid{nppp}

\jnltitlelogo{Nuclear and Particle Physics Proceedings}

\usepackage{amssymb}
\usepackage[fleqn]{amsmath}
\usepackage{siunitx}
\newcommand{\gev}[1]{\SI{#1}{\giga\electronvolt}}
\newcommand{\tev}[1]{\SI{#1}{\tera\electronvolt}}

\usepackage[figuresright]{rotating}
\usepackage{subcaption}

\newcommand{\hc}{\text{H}}
\newcommand{\dd}{\mathrm{d}}

\newcommand{\Eq}[1]{Eq.~(\ref{#1})}

\newcommand{\Fig}[1]{Fig.~\ref{#1}}

\newcommand{\Eqs}[1]{Eqs.~(\ref{#1})}

\begin{document}

\begin{frontmatter}

\title{Small-$p_T$ and large-$x$ regions for Higgs transverse momentum distributions}

\author[label1]{Tanjona R. Rabemananjara}
\ead{tanjona.rabemananjara@mi.infn.it}
\address[label1]{Tif Lab, Dipartimento di Fisica, Universita di Milano, and \\
INFN, Sezione di Milano, Via Celoria 16, I-20133 Milano, Italy}

\pagestyle{myheadings}
\markright{ }

\begin{abstract}
It was shown recently that standard resummation of logarithms of $Q/p_T$ can be supplemented 
with the resummation of logarithmic contributions at large $x=Q^2/s$ in the case of a colourless
final state such as Higgs produced via gluon fusion or the production of a lepton pair via 
Drell--Yan mechanism. Such an improved transverse momentum resummation takes into account soft
emissions that are emitted at very small angles. We report on recent phenomenological studies of
a combined threshold-improved $p_T$ and threshold resummation formalism to the Higgs boson
produced at the LHC where small-$p_T$ and threshold logarithms are resummed up to NNLL and NNLL*
respectively. We show that the effect of the modified $p_T$ resummation yields a faster perturbative
convergence in the small-$p_T$ region while the effect of the threshold one improves the agreement 
with fixed-order calculations in the medium and large-$p_T$ regions.
\end{abstract}


\end{frontmatter}
\section{Introduction}
\label{sec:introduction}

Precise computations of differential distributions of a colour singlet object such as the Higgs boson will
play a major role in the future physics programme of the LHC. In particular, the kinematic distributions of
the Higgs boson produced in association with QCD radiations can be used to constrain new-physics models
due to its sensitivity to modification of the Yukawa couplings \cite{Soreq:2016rae, Bishara:2016jga}. The 
transverse momentum distribution of the Higgs in the Effective Field Theory (EFT) approach is currently known 
to very high accuracy, namely NNLO \cite{Boughezal:2015aha, Caola:2015wna, Chen:2016zka}. The perturbative description of the Higgs transverse momentum in certain regions of the phase space, however, requires the resummation of large logarithmic corrections to all-order in the coupling constant $\alpha_s$.

In this article, we review a formalism \cite{Muselli:2017bad, Rabemananjara:2020rvw} that combines two regions,
namely small-$p_T$ and large-$x$, for the resummation of Higgs transverse momentum spectra. This was achieved by, 
first modifying the standard $p_T$ resummation to take into account soft radiations hat emitted at very small 
angles, and then combining the resulting expression with the pure contribution from the threshold. 

Our starting point is the Mellin space cross section \cite{Muselli:2017bad, Rabemananjara:2020rvw} where short and large range interactions factorizes:
\begin{align}
\frac{\mathrm{d} \sigma^{\text{res}}}{\mathrm{d} \xi_p} (N, \xi_p, \alpha_s) = \sum_{a,b} \mathcal{L}_{ab}(N)  
\frac{\mathrm{d} \hat{\sigma}^{\text{res}}_{ab}}{\mathrm{d} \xi_p} (N, \xi_p, \alpha_s),
\label{eq:Mellin}
\end{align}
where we have expressed the cross sections in terms of the dimensionless variable $\xi_p = p_T^2 / Q^2$.
In the sequel, the superscript "res" may describe the threshold, modified small-$p_T$, and combined resummation.
This article starts with a brief description of the three resummation formalisms, followed by the phenomenological
results for the Higgs boson produced at LHC. Conclusions are drawn in Section~\ref{sec:conclusions}.

\section{Threshold resummation}
\label{sec:threshold}

This section is not meant to be a review of the threshold resummation for Higgs boson production, in which
extensive literature~\cite{deFlorian:2005fzc} already exists,  but rather to collects the relevant equations 
that will be used as a comparison to the threshold-improved $p_T$ resummation in Section~\ref{sec:tipt}.

Typically, when the invariant mass $Q$ of the system approaches the partonic center of mass energy $\sqrt{\hat{s}}$,
the phase space for gluon bremsstrahlung vanishes and results in large logarithmic corrections. The tower
of logarithms are of the form $\alpha_s^n \left( \ln^k (1-x)/(1-x) \right)_{+}$ for $0 \leq k \leq 2n-1$
at the $k$-th order in perturbation theory. Here, the plus distribution regularizes the singularity at
$x = 1$. The appearance of these large corrections spoils the perturbative convergence of the fixed-order
calculations even if we are well in the perturbative regime where $\alpha_s \ll 1$. 
These logarithmic structures, however, exponentiate after taking the Mellin transform w.r.t. the relevant
kinematic variable. Generally, threshold resummed expression for the production of a Higgs boson takes the form
\begingroup
\allowdisplaybreaks
\begin{align}
\frac{\dd \hat{\sigma}_{ab}}{\dd \xi_p} & \left( N, \xi_p, \alpha_s \right) = 
\mathcal{C}_{ab}^{\text{(LO)}} \left( N, \xi_p, \alpha_s \right) g_{ab} \left( \xi_p, \alpha_s \right)  
\nonumber \\
& \exp \left( \Delta_i(N) + \Delta_j (N) + J(N) + S (N, \xi_p) \right),
\end{align}
\endgroup 
where $\sigma_0$ denotes the leading order Born cross section. Notice that we have omitted the $\alpha_s$
dependence in the exponent for simplicity. The matching function $g$ assures that at
every perturbative order, the resummed cross section agrees with the exact fixed-order calculations up to
corrections of the order of $1/N$. The exponent collects all the enhanced terms when $N \to \infty$ 
(equivalent to say $x \to 1$) which in itself can be written as perturbative series $\sum_{n=0}^{\infty}
\sum_{l=1}^{n+1} a_{n,l} \alpha_s^n \ln^l (N)$ . Notice that now, the enhanced contributions are
single-logarithmic in the moment $N$. Finally, $\mathcal{C}^{\text{(LO)}}$ represents the leading-order term
in the $p_T$ distribution whose $gg$-channel part is given by
\begingroup
\allowdisplaybreaks
\begin{align}
\mathcal{C}_{gg}^{\text{(LO)}} & \left( N, \xi_p, \alpha_s \right) = 2 \alpha_s \frac{\sigma_0}{\xi_p} \frac{C_A}{\pi} 
\sum_{m=0}^{4} \left( - 1 \right)^m \frac{\Gamma(1/2) \Gamma(N+m)}{\Gamma(N+m+1/2)} \nonumber \\
& \hspace*{0.75cm} f_m (\xi_p) \, _2F_1 \left( \frac{1}{2}, N+m, N+m+\frac{1}{2}, z^2 \right).
\label{eq:lo-pt}
\end{align}
\endgroup
The functions $f_m$ fully depend on the modified variable $\xi_p$ and are defined as follows:
\begingroup
\allowdisplaybreaks
\begin{flalign}
f_0 & (\xi_p) = 1, &f_2 & (\xi_p) = z^2 (1+\xi_p)(3+\xi_p) \nonumber \\
f_1 & (\xi_p) = z (1+\xi_p), &f_3 & (\xi_p) = z^3 (1+\xi_p) \\
f_4 & (\xi_p) = z^4, \text{ with } &z& \equiv z(\xi_p) = \left( \sqrt{1+\xi_p} + \sqrt{\xi_p} \right)^{-2}. \nonumber
\end{flalign}
\endgroup

\section{Threshold-improved $p_T$ resummation}
\label{sec:tipt}

\subsection{Description}
\label{subsec:description}

In the context of threshold-improved $p_T$ resummation~\cite{Muselli:2017bad, 
Rabemananjara:2020rvw}, soft and small-$p_T$ behaviours are
jointly resummed in the new argument of the logarithm $\chi = \left( \bar{N}^2 + \hat{b}^2 /b_0^2 \right)$
where $\bar{N} = N \exp(\gamma_E)$ and $\hat{b} = bQ$.
It is now apparent that $\chi$ interpolates between threshold and small-$p_T$ limit in the respective
limit. As shown in \cite[Eq.~(2.7)]{Rabemananjara:2020rvw}, the partonic $(N, \xi_p)$-space of the 
hadronic resummed cross section is given by
\begin{align}
\frac{\dd \sigma_{ab}^{\text{tr'}}}{\dd \xi_p} ( \xi_p ) =  \frac{\sigma_0}{z(\xi_p)^N}
\int_{0}^{\infty} \dd \hat{b} \frac{\hat{b}}{2} J_0 (\hat{b} \sqrt{\xi_p}) 
\frac{\dd \hat{\sigma}_{ab}^{\text{tr'}}}{\dd \xi_p} \left( b \right),
\label{eq:fill-tipt}
\end{align}
where the explicit expression of the $(N,b)$-space partonic part is given by the following:
\begin{flalign}
\frac{\dd \hat{\sigma}_{ab}^{\text{tr'}}}{\dd \xi_p} & \left( N, b, \alpha_s \right) = 
\bar{\hc}_g \left( \frac{\bar{N}^2}{\chi}, \alpha_s (Q^2) \right) 
\mathcal{W}_{ga} \left(N, \alpha_s\left( \frac{Q^2}{\chi} \right) \right) \nonumber \\
& \mathcal{W}_{gb} \left(N, \alpha_s\left( \frac{Q^2}{\chi} \right) \right) 
\exp \left( S_g \left( N, \chi, \alpha_s (\mu_R^2) \right) \right).
\label{eq:tipt}
\end{flalign}
The hard function $\bar{\mathrm{H}}$ collects all the $b$-depedence that are not enhanced in the limit
$b \to \infty$. Its expression can be written in terms of the standard hard function whose perturbative
expansions are given in \cite{Catani:2011kr},
\begin{align}
\bar{\hc}_g \left( \frac{\bar{N}^2}{\chi}, \alpha_s \right) = \mathrm{H}_g (\alpha_s) + A_1 \alpha_s
\mathrm{Li}_2 \left( \frac{\bar{N}^2}{\chi} \right) + \mathcal{O}(\alpha_s^3).
\end{align}
The function $\mathcal{W}$ contains both the coefficient function and the evolution of the Parton 
Distribution Functions (PDFs). Its expression is given by
\begin{align}
\mathcal{W}_{ga} \left(N, \alpha_s\left( \frac{Q^2}{\chi} \right) \right) = C_{gi} \left(N,
\alpha_s\left( \frac{Q^2}{\chi} \right) \right) U_{ia} \left( N, \alpha_s\left( \frac{Q^2}{\chi}
\right) \right)
\end{align}
where the expressions of $C$ and $U$ are defined as in~\cite[Eq.~(2.26), Eq.~(2.28)]{Rabemananjara:2020rvw}.
As opposed to the standard $p_T$ resummation \cite{Bozzi:2005wk}, the coefficient and evolution functions are 
computed from a joint scale $Q^2 / \chi$. The logarithmic enhanced contributions are contained in $S$ and is 
defined such that $S=1$ when  $\alpha_s \ln \chi = 0$. The LL term collects logarithmic contributions of the 
form $\alpha_s^n \ln^{n+1} \chi$; the NLL part resums logarithms of the form $\alpha_s^n \ln^{n} \chi$; and
finally the NNLL terms resums $\alpha_s^n \ln^{n-1} \chi$ contributions. Therefore, the function $S$ can
be organized as follows:
\begin{align}
S_g (N, \chi, \mu_R^2) = \alpha_s^{-1} g_1(\chi) + g_2(\chi) + \alpha_s g_3(\chi),
\end{align}
where
\begingroup
\allowdisplaybreaks
\begin{align}
g_{1} (\chi) &= \frac{A_1}{\beta_{0}^{2}} (\lambda_{\chi}+\ln (1-\lambda_{\chi})) \:
\text{with} \: A_1 = \frac{C_A}{\pi} .
\end{align}
\endgroup
Here, we only give the expression of the leading logarithmic contributions that will be used later. The explicit 
expressions of $g_1$ and $g_2$ can be extracted from \cite{Muselli:2017bad}.
The $b$ and $N$ dependence are now embodied in $\lambda_{\chi} = \alpha_s \beta_{0} \ln \chi$ while the pure
$N$ dependence is embodied in $\lambda_N = \alpha_s \beta_{0} \ln \bar{N}^2$. It is straightforward to see
that the in the small-$p_T$ limit (or equivalently large-$b$), threshold-improved $p_T$ reproduces standard 
transverse momentum as given in Ref.~\cite{Bozzi:2005wk}. Indeed, taking $b \to \infty$ in~\Eq{eq:tipt}
implies that $Q^2 / \chi \to b_0^2 / b^2$, $\chi \to \hat{b}^2 / b_0^2$, and $\mathrm{Li}_2 (\bar{N}^2 / \chi) \to 0$.

\subsection{Large-$N$ behaviour at small-$p_T$}
\label{subsec:lnpt}

In the following section, we check that the threshold-improved $p_T$ resummation captures the correct threshold
behaviour at small-$p_T$; these are initial state soft gluons emitted at very small angles. Such a check amounts
to the expansion of the resummed expression in \Eq{eq:fill-tipt} and compare it to $\mathcal{C}^{({LO)}}$ in 
\Eq{eq:lo-pt}. For simplicity, we only consider the LL resummation of the $gg$-channel with the $\alpha_s$-term 
in the perturbative expansion.

\begin{figure}[!h]
\includegraphics[width=\linewidth]{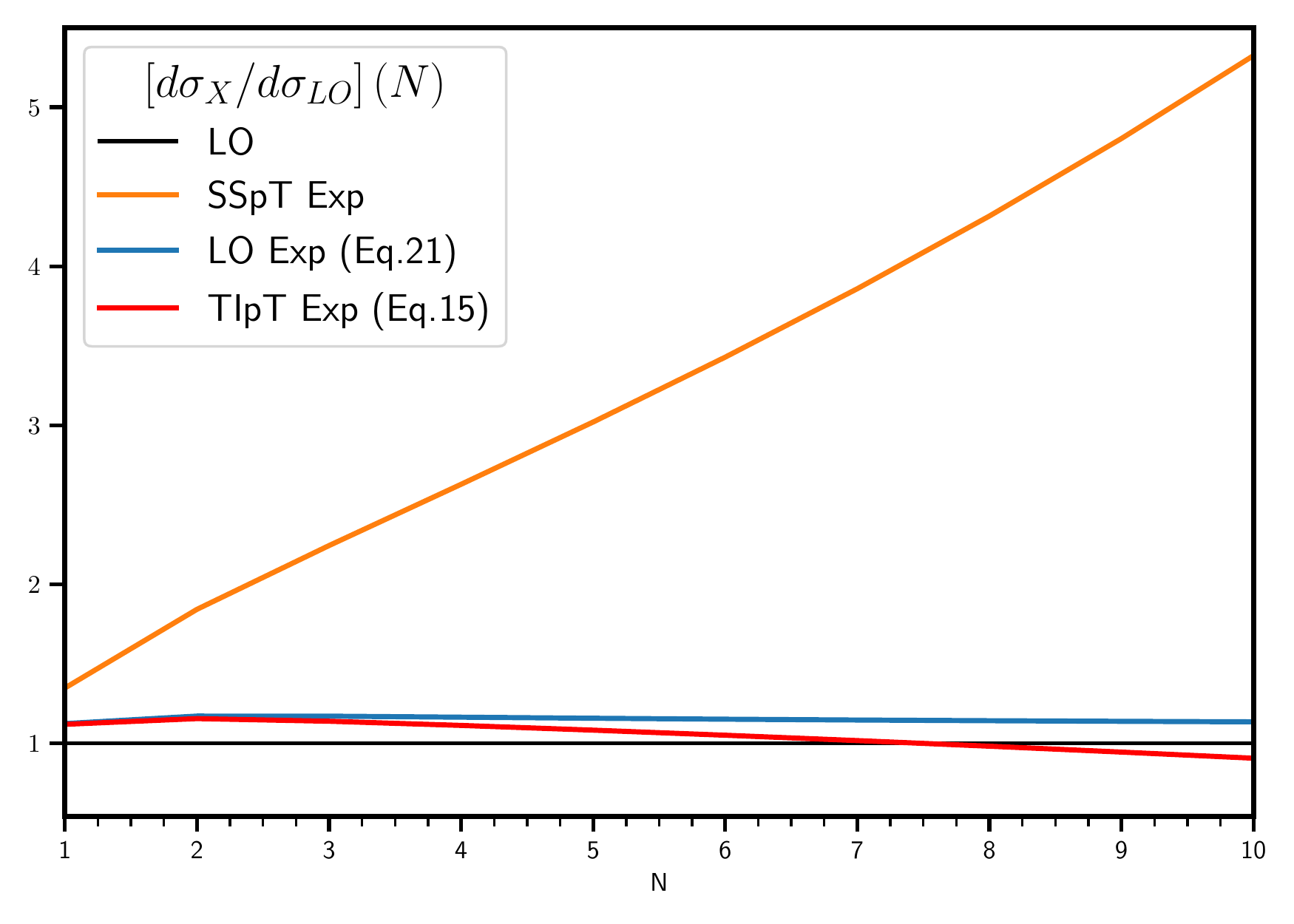}
\caption{Ratio between the asymptotics and the LO in \Eq{eq:lo-pt}. The Standard $p_T$ (SSpT)
resummation is computed in the limit $\xi_p \to 0$. In the plots, $\alpha_s = 0.118$ and $\xi_p = 10^{-3}$.}
\label{fig:asymp}
\end{figure}

We start from \Eq{eq:tipt} where at LL we approximate $\bar{\mathrm{H}}_g$ and the coefficient function $C_{gg}$ to 1
and we only include in the Sudakov exponent the function $g_1$. The PDF evolutions are included up to LO with the
leading-order anomalous dimension replaced by its large-$N$ behaviour while the evolution factor that evolves the
coefficient functions are approximated to $1$. Thus, using \cite[Eq.~(2.28a)]{Rabemananjara:2020rvw}, we arrive to
the following
\begin{align}
\frac{\dd \hat{\sigma}_{gg}^{\text{tr'}}}{\dd \xi_p} \left( N, b, \alpha_s \right) = \left( U^{\text{(LO)}}_{gg,N} 
\left( \frac{Q^2}{\chi} \leftarrow Q^2 \right) \right)^2 \exp\left( g_1 \right)
\label{eq:ll_lo}
\end{align}
The LO solution \cite[Eq.~(2.29)]{Rabemananjara:2020rvw} to the DGLAP evolution equation in the large-$N$ limit
writes as
\begin{align}
U^{\text{(LO)}}_{gg,N} \left( \frac{Q^2}{\chi} \leftarrow Q^2 \right) = - \frac{A_1}{2\alpha_s\beta_0^2} 
\lambda_N \ln(1-\lambda_\chi) .
\label{lowest-order-evolution}
\end{align}
Thus, \Eq{eq:ll_lo} yields
\begin{align}
\frac{\dd \hat{\sigma}_{gg}^{\text{tr'}}}{\dd \xi_p} \left( N, b, \alpha_s \right) = \exp \left( \frac{A_1}
{\alpha_s \beta_{0}^2} \left( \lambda_{\chi} + (1 - \lambda_N) \ln (1-\lambda_{\chi}) \right)  \right)
\end{align}
Expanding the above equation and retaining only the $\alpha_s$-term we arrive at
the following expression
\begin{align}
\frac{\dd \hat{\sigma}_{ab}^{\text{tr'}}}{\dd \xi_p} \left( N, b, \alpha_s \right) = - \alpha_s A_1 \left( 
\frac{\ln^2 \chi}{2} - 2 \ln \bar{N} \ln \chi \right) .
\label{eq:exp-1}
\end{align}
Plugging back \Eq{eq:exp-1} into \Eq{eq:fill-tipt} and performing the Fourier back transform using
\cite[Eq.~(2.43a), Eq.~(2.43b)]{Rabemananjara:2020rvw} we end up with a $(N,\xi_p)$ version of the
expanded expression. The final result writes as,
\begin{align}
\frac{\dd \sigma_{gg}^{\text{tr'}}}{\dd \xi_p} ( N, & \xi_p, \alpha_s ) = 2 \alpha_s A_1 \frac{\sigma_0}{z(\xi_p)^N}
\left\lbrace \frac{N}{\sqrt{\xi_p}} \mathrm{K}_1^{(1)} (2 N \sqrt{\xi_p}) \, \nonumber \right. \\
- & \left. \frac{N}{\sqrt{\xi_p}} \left( \ln (N \sqrt{\xi_p})  + \gamma_E \right) \mathrm{K}_1 (2 N \sqrt{\xi_p}) 
\right\rbrace .
\label{eq:exp_ll_lo}
\end{align}
Recall that we are here interested to check the large-$N$ behaviour of the threshold-improved transverse
momentum resummation while keeping $N p_T$ fixed in which case the expression above cannot be simplified further.
Thus, in order to relate \Eq{eq:exp_ll_lo} with \Eq{eq:lo-pt}, we have to simplify $\mathcal{C}_{gg}^{\text{(LO)}}$
in the aforementioned limit. For this, it is sufficient to only consider the most dominant term, i.e. $m=4$.
Using the properties of the Gauss hypergeometric functions, we have
\begin{align}
_2F_1 \left( \frac{1}{2}, N + 4, N + \frac{9}{2}, z^2 \right) = \frac{_2\tilde{F}_1 \left(N + \frac{9}{2};
\frac{z^2}{z^2 - 1} \right)}{\sqrt{1 - z^2}},
\label{eq:hyp-def}
\end{align}
where we have defined $_2\tilde{F}_1 \left(N+9/2;z^2 / (1-z^2) \right) = \, _2F_1 \left( 1/2, 1/2 N+9/2;z^2(1-z^2) 
\right)$ for simplicity. In the limit we are interested in, $_2F_1$ can be simplified further. Using results from
Refs.~\cite{FaridKhwaja2014,Temme2003},
\begin{align}
_2\tilde{F}_1 & \left(N + \frac{9}{2}; \frac{z^2}{z^2 - 1} \right) = \frac{\Gamma(N+9/2)}{\Gamma(N+4)} 
\sum_{s = 0}^{\infty} g_s \left( \frac{z^2}{z^2 - 1} \right) \nonumber \\
& \frac{\Gamma(s + 1/2)}{\Gamma(1/2)} \ln^s \frac{1}{z^2} \mathrm{U} \left( s + \frac{1}{2}, s + 1,
- 2 N \ln z \right).
\label{eq:expand_2f1}
\end{align}
We stress that this expression only holds if $N \to \infty$ uniformly w.r.t. large values of $z^2 / (z^2 - 1)$
which is indeed the case as $z^2 / (z^2 - 1) \to \infty$ as $\xi_p \to 0$. In such a limit, it is sufficient to
only consider first term in the sum of \Eq{eq:expand_2f1} (i.e. $s=0$) where $g_0$ and the 2nd Kummer function
$\mathrm{U}$ are defined as
\begin{align}
& g_0  \left( z \right) = \sqrt{\ln \left( \frac{z - 1}{z} \right)}, \label{eq1}\\
& \mathrm{U} \left( \frac{1}{2}, 1, 2 z \right) =  \frac{1}{\Gamma(1/2)} z \exp(z) \,
\mathrm{K}_1^{(1)} (z) . \label{eq2}
\end{align}
Using \Eqs{eq1} and~(\ref{eq2}) to redefine $_2F_1$ in \Eq{eq:hyp-def} and plugging its expression back into
\Eq{eq:lo-pt} we arrive to the following expression
\begingroup
\allowdisplaybreaks
\begin{align}
\left. \mathcal{C}_{gg}^{\text{(LO)}} \right\rvert_{ m=4} = 2 \alpha_s \frac{\sigma_0}{\xi_p} \frac{C_A}{\pi} 
\frac{ \tilde{z} \sqrt{2 \ln z} }{z^{N+2} \sqrt{1 - z^2}} \mathrm{K}_1^{(1)} (\tilde{z}),
\end{align}
\endgroup
with $\tilde{z} = -N \ln z$. The $z^N$-term in the denominator comes from re-rewriting the exponential in 
\Eq{eq2} in terms of the explicit expression of $z$. We can now safely take the limit $\xi_p \to 0$ in the terms
that are not $N$-dependent, which yields the following simplifications $\sqrt{2 \ln z}\sim 2 \sqrt[4]{\xi_p}$ and 
$\sqrt{1-z^2} \sim 2 \sqrt[4]{\xi_p}$. Finally, the large-$N$ limit just amounts to replacing $N+2$ with $N$. 
Putting everything
\begingroup
\allowdisplaybreaks
together and with a little bit of algebraic simplification we get
\begin{align}
\left. \mathcal{C}_{gg}^{\text{(LO)}}  \right\rvert_{ m=4} = 4 \alpha_s \frac{C_A}{\pi} \frac{\sigma_0}{z^N}
\frac{N}{\sqrt{\xi_p}} \mathrm{K}_1^{(1)} (2 N \sqrt{\xi_p}) .
\end{align}
\endgroup
In the limit $N \sqrt{\xi_p} \to 0$, the above expression is \emph{asymptotic} to expansion of the modified resummation
in \Eq{eq:exp_ll_lo} as shown in \Fig{fig:asymp}. In order to generate $\mathrm{K}_1$-terms from the leading-order $p_T$ 
spectra $C_{gg}^{\text{(LO)}}$, the sum in \Eq{eq:expand_2f1} has to be performed beyond $s=0$. This is computationally challenging as the form of the Kummer $\mathrm{U}$ does not lead directly to compact Bessel functions. However, it can be seen from the numerical checks in \Fig{fig:asymp} that the threshold-improved
$p_T$ resummation captures best the soft behaviours from fixed-order calculations while the standard $p_T$
resummation deviates largely from the LO as $N$ increases.

\section{Combined resummed expression}
\label{sec:combined}

The threshold-improved $p_T$ resummation described in Section~\ref{sec:tipt} does not contain all the soft
logarithms such as the soft logarithms that arise when soft gluons are emitted at large angle. Therefore,
the pure threshold contribution has to be manually added through a profile matching function. The profile
matching function is chosen such (i) that there is no double counting, and (ii) the combined result
reproduces threshold-improved $p_T$ and threshold resummation at small $p_T$ and large $x$ respectively. A
possible expression for such a combined resummation is given by
\begin{align}
\frac{\mathrm{d} \hat{\sigma}_{ab}}{\mathrm{d} \xi_p} (N,\xi_p, \alpha_s) &= \mathrm{T} (N, \xi_p) 
\frac{\mathrm{d} \hat{\sigma}_{ab}^{\text{thrs}}}{\mathrm{d} \xi_p} (N, \xi_p, \alpha_s) \: + \nonumber \\
& \left( 1- \mathrm{T} (N, \xi_p) \right) \frac{\mathrm{d}
\hat{\sigma}_{ab}^{\text{tr'}}}{\mathrm{d} \xi_p} (N, \xi_p, \alpha_s),
\label{eq:combined_expr}
\end{align}
where we define the profile matching function as $\mathrm{T} (N, \xi_p) = N^k \xi_p^m / (1 + N^k \xi_p^m)$.
Hence, combined resummation produces results that differ from the threshold-improved $p_T$ resummation 
by $\mathcal{O}(\xi_p^m)$ corrections when $\xi_p \to 0$, and from the threshold resummation by
$\mathcal{O} (1/N^k)$ when $N \to \infty$. The values of $k$ and $m$ can be chosen arbitrarily provided
that $m < k$ and can be used to assess the ambiguity of the matching.

\section{Phenomenological results}
\label{sec:analyses}

In this section, we present phenomenological results for the standard model Higgs produced via gluon-gluon
fusion at the LHC with a center of mass energy $\sqrt{s}=\tev{13}$. The issue related to the computation of
the Fourier-Mellin back-transform from our modified $p_T$ resummation is dealt using the Borel prescription~\cite{Rabemananjara:2020rvw}. The
following plots are produced using the  \textbf{\texttt{NNPDF31$\_$nnlo$\_$as$\_$0118}} set~\cite{Ball:2014uwa}
of parton distribution functions. The uncertainty bands are obtained from varying the renormalization and 
factorization scales using the seven-point prescription. For the combined results, the parameters that enter 
into the matching function are set to $m=2$ and $k=3$.
\begin{figure}[!h]
\captionsetup[subfigure]{aboveskip=-1.5pt,belowskip=-1.5pt} 
\centering
\begin{subfigure}{0.95\linewidth}
\includegraphics[width=\linewidth]{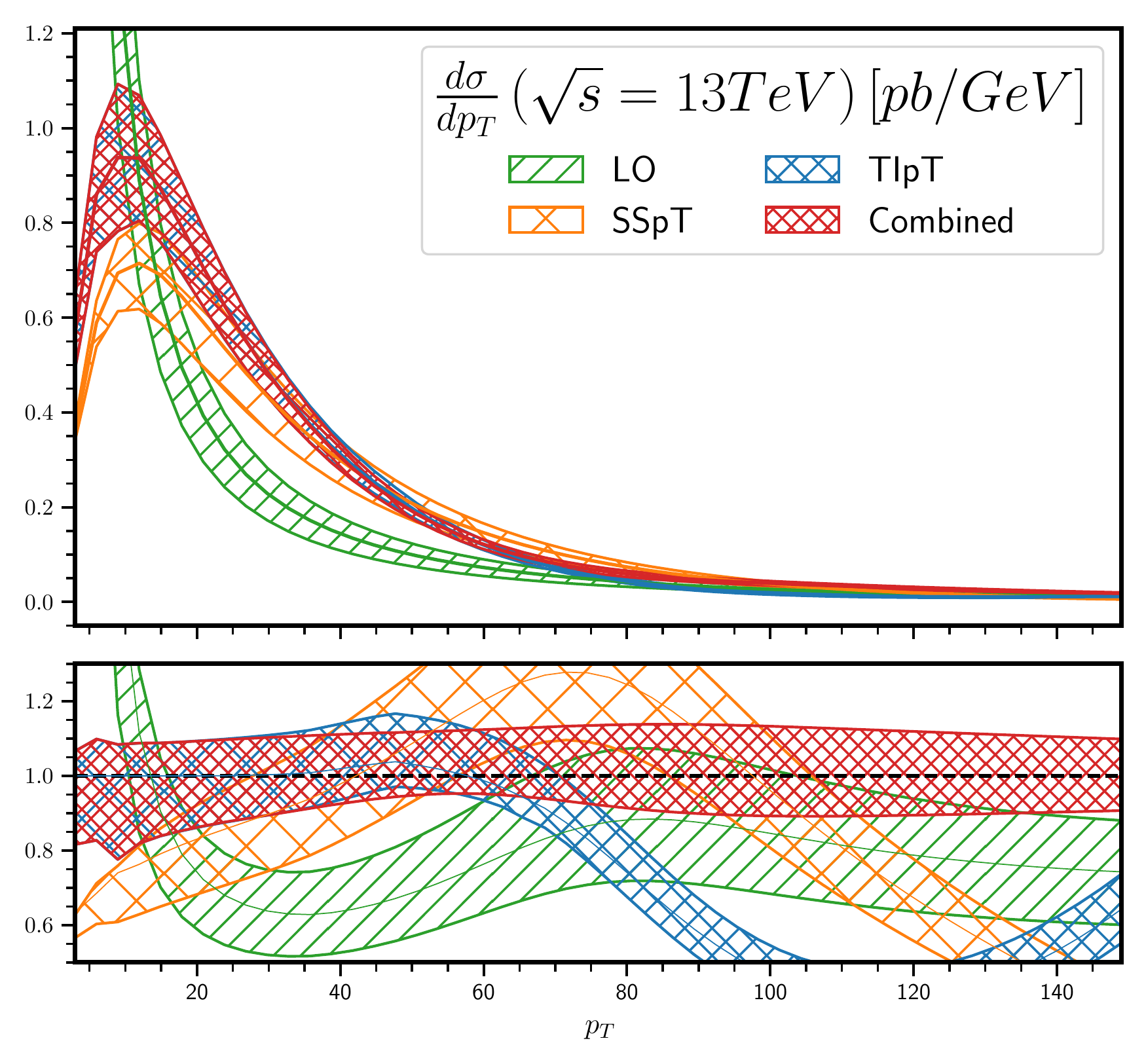}
\caption{NLL+LO} 
\label{higgs:sspt-matched}
\end{subfigure}
\begin{subfigure}{0.95\linewidth}
\includegraphics[width=\linewidth]{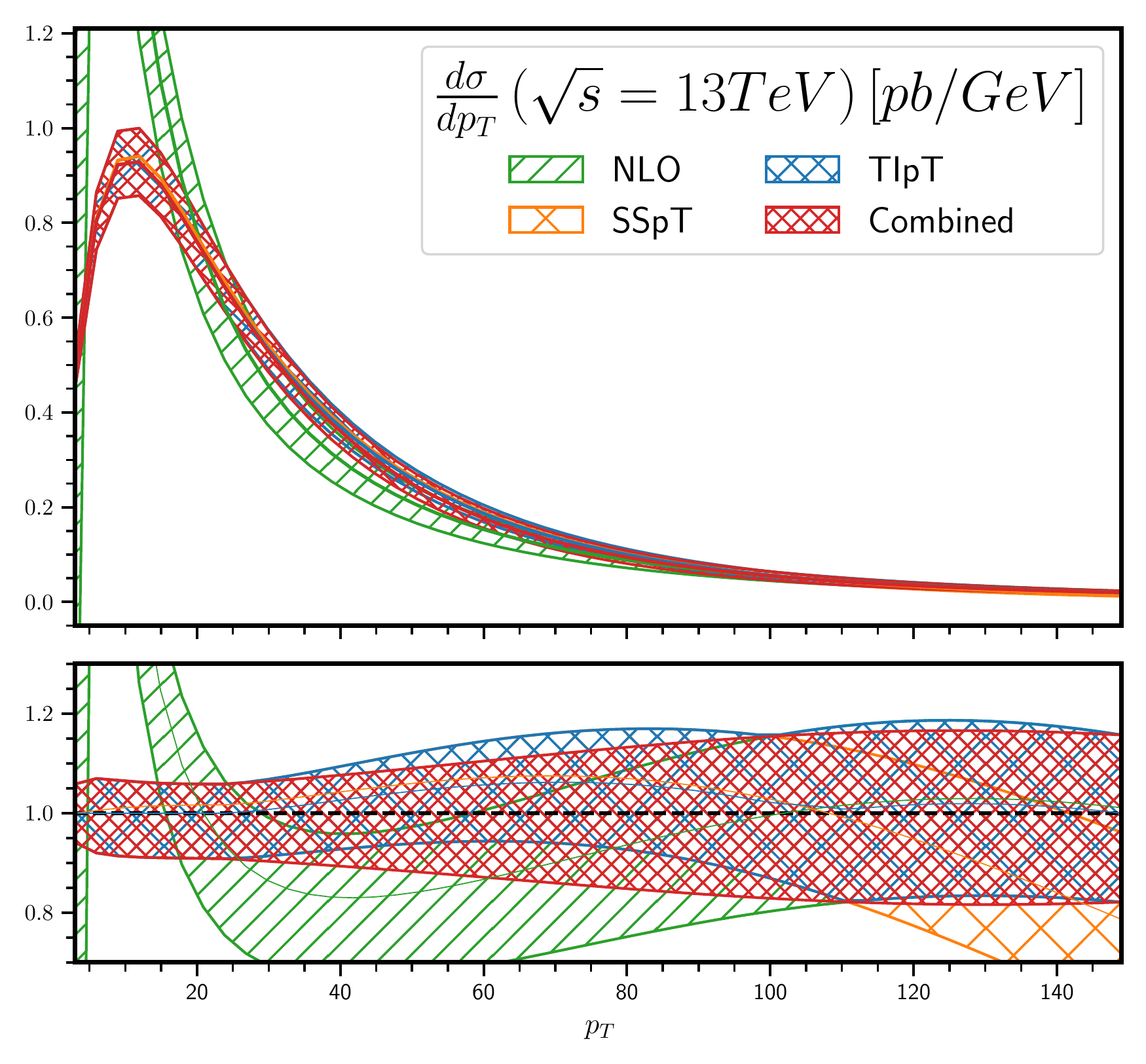}
\caption{NNLL+NLO}
\label{higgs:consistent-matched} 
\end{subfigure}
\caption{$p_T$ spectrum for the Higgs boson production comparing the three types of resummations matched
with standard fixed-order calculations. The top panels compare the matched results with the F.O calculations, 
while the lower panels show the ratio w.r.t. the central value of the combined results.}
\label{fig:higgs-pheno} 
\end{figure}

In \Fig{fig:higgs-pheno}, we compare the standard $p_T$ (SSpT) resummation with the threshold-improved $p_T$
(TIpT) and combined resummation. The resummed calculations are matched with the fixed-order (F.O) predictions
in order to reduce the effect of unjustified logarithms in the medium and large-$p_T$ regions. In contrast to
the NLO predictions, resummation leads to a well-behaved transverse momentum distribution in the small-$p_T$
regions that has a
peak at $p_T \sim \gev{10}$. As we go from SSpT to TIpT, we see a faster perturbative converge of the NLL+LO
result while the NNLL+NLO results are similar. This is well understood as the $N/b$ corrections become
negligible as the logarithmic accuracy increases. Due to the structure of the matching function, TIpT and
combined are exactly similar at small-$p_T$ and the pure threshold contribution start to become relevant
at $p_T \sim \gev{60}$ which yields a better agreement with the F.O results. This is because medium and 
large-$p_T$ regions are dominated by soft behaviours.

\section{Conclusions}
\label{sec:conclusions}

In this article, we have presented a review of a formalism that combines small-$p_T$ and large-$x$ resummation
for Higgs boson transverse momentum distribution. Such a formalism results in resummation of transverse
momentum distributions that is valid for all values of momenta. In particular, we investigated further the
soft logarithms that are present in the modified $p_T$ resummation by comparing it to fixed-order. We finally
showed that while the modified resummation improves the convergence at small-$p_T$, combining it with the pure
threshold contribution results in a more accurate prediction in the medium and large-$p_T$ regions.

\vspace*{-0.35cm}
\nocite{*}
\bibliographystyle{elsarticle-num}
\bibliography{refs.bib}

\end{document}